# Inverse-Gaussian-Apodized Fiber Bragg Grating for Dual Wavelength Lasing

**Bo Lin[1], Han Zhang[1], Swee Chuan Tjin[1,*], Dingyuan Tang[1], Jianzhong Hao[2], Chia Meng Tay[1] and Sheng Liang[3]**

[1]*Photonics Research Centre, School of Electrical and Electronic Engineering, Nanyang Technological University, 50 Nanyang Avenue, Singapore 639798, Singapore*

[2]*Institute for Infocomm Research, 1 Fusionopolis Way, Singapore 138632, Singapore*

[3]*School of Instrument Science and Opto-electronics Engineering, Beihang University, 37 Xueyuan Road, Beijing 100191, China*

*Corresponding author: esctjin@ntu.edu.sg

A fiber Bragg grating (FBG) with an inverse-Gaussian apodization function is proposed and fabricated. It is shown that such a FBG possesses easily controllable dual-wavelength narrow transmission peaks. Incorporating such a FBG filter in a fiber laser with a linear cavity, stable dual-wavelength emission with 0.146 nm wavelength spacing is obtained. It provides a simple and low cost approach of achieving the dual-wavelength fiber laser operation.

## 1. Introduction

Fiber Bragg gratings (FBGs) are an important optical component widely used in fiber optical communication systems [1,2] and fiber lasers [3,4]. In fiber lasers, the versatility of FBGs allows them to function as selective wavelength filters as well as broadband reflectors in a Fabry-Perot fiber cavity. A FBG is generated by periodically modulating the refractive index of the fiber core. It is well-known that a uniform FBG has large sidelobes, which can nevertheless be suppressed through apodizing the uniform FBG with a Gaussian function [5]. So far, apodized FBGs with various apodization functions have been extensively investigated, however, no work presented to enhance and thus to make use of the unwanted sidelobes, to the best of our knowledge.

Different types of FBGs were used as a dual-channel filter in fiber lasers previously [6-12]. However, each of the reported works has limitations. For example, a nanometer precision translation stage (thus high cost) was required in [6] and [7]. The dual-channel lasers described in [8-10] used polarization maintaining (PM) fibers, which are more expensive than the standard single-mode fiber. In [11] a FBG pair was used as a dual-channel filter, it cannot be fabricated in a single process when the length of the FBG pair (including the spacing between the two separated FBGs) is larger than the phase mask length. Chen et al. [12] proposed an equivalent phase shift (EPS) FBG as a filter in the ring laser cavity. Although only relatively cheap translation stage with micrometer precision is needed to fabricate the FBG, the sampling period of the grating in the experiment needs to be adjusted when compared to the calculated result.

To overcome the above-mentioned limitations, in this paper, we propose a new FBG structure. We show that inversely apodizing a FBG with an inverse-Gaussian function the sidelobes of the FBG could be greatly enhanced, consequently controllable dual-wavelength transmission peaks are formed in the transmission spectrum of the FBG. Such an inverse-Gaussian apodized FBG (IGAFBG) can be used as a flexible dual-wavelength bandpass filter. Incorporating the FBG filter in a linear cavity fiber laser we have achieved stable dual-wavelength emission with 0.146 nm wavelength separation. The proposed technique is a single-step fabrication process which is flexible to change the bandwidths, separation of the two pass bands through changing the design parameters of the IGAFBG, such as the grating length, index modulation depth, and inverse apodization function type.

## 2. Principle of the IGAFBG and Simulation Results

Using the transfer matrix method [5], we have simulated the reflection spectrum of a proposed IGAFBG, as shown in Fig. 1(a). In calculation, the following parameters were used: the effective refractive index $n_{eff}$ is 1.447, the grating length $L$ is 12 mm, the grating period $\Lambda$ is 532.85 nm, the modulation depth $\overline{\delta n_{eff}}$ is $3\times10^{-4}$ and the inverse apodization function used is $A(z) = 1 - \exp\left\{\left[-4(\ln 2)z^2\right]/(L/3)^2\right\}$ (the grating starts from -$L$/2 to $L$/2 in the $z$ dimension). Fig. 1(b) shows the effective index variation along the fiber axis of an IGAFBG, in which the average index variation (dashed line) has an inverse-Gaussian distribution. For illustrative purposes, the size of the grating period is greatly exaggerated relative to the grating length. A traditional Gaussian apodized FBG is also simulated by using the same parameters of the IGAFBG except that the apodization function is $B(z) = \exp\left\{\left[-4(\ln 2)z^2\right]/(L/3)^2\right\}$, as shown in Fig. 2. In a traditional apodized grating [5, 13, 14] the average refractive index is not uniform along the length of the grating. There are sidelobes appearing on the short-wavelength side of Bragg wavelength, however, no sidelobes are observed on the long-wavelength side, as shown in Fig. 2. It was due to the non-uniform "dc" index change caused by the Fabry-Perot effect [15]. Singh et al. [16] explained that the suppression of sidelobes was not pronounced because the self-induced chirp due to the non-uniform grating makes the resonant wavelengths on both ends of the grating smaller than the resonant wavelengths of the central part of the grating. Therefore, the sidelobes only exist on the short-wavelength side of the spectrum. The inverse-Gaussian apodization function used in our work is unity minus the commonly used Gaussian apodization function. <u>The resonant wavelengths on both ends of the grating are larger than the resonant wavelength in the central portion of the grating and thus the sidelobes on the long-wavelength side of the spectrum are enhanced, and no sidelobe enhancement occurs on the short-wavelength side of Bragg wavelength</u>. The enhancement of the sidelobes can be raised to such a level that the gaps between the lobes are deep and narrow enough to form high transmission pass bands, as clearly shown in Fig. 1(a).

## 3. Experimental Results of the Proposed IGAFBG

To verify the simulation result, we fabricated an IGAFBG with a uniform phase mask based on the phase-mask scanning technique. The UV light (70 mW) from a 248-nm frequency-doubled Argon laser is focused through a cylindrical lens and a uniform phase mask with a pitch of 1065.7 nm, onto the core of a single-mode fiber. The fiber is hydrogen-loaded for 5 days. The inverse-Gaussian apodization function is realized by varying the scanning speed of the laser beam. A high scanning speed yields a low index modulation depth (low UV light exposure). Through adjusting the laser beam power, beam size, the position of the fiber relative to the phase mask, different index modulation depths could be achieved. It is to note that the type of photosensitive fiber and the amount of hydrogen loading into the fiber affect the effective refractive index. After several trials of different scanning speed distributions, we chose the scanning speed profile $V(z) = 0.03 \times (A(z) + 0.1)^{-1}$ mm/s to best fit the inverse-Gaussian apodization function *A(z)*. An IGAFBG with a length of 1.2 cm was fabricated. The experimentally measured (solid line) and the numerically calculated (dashed line) transmission spectra are shown in Fig. 3. Two pass bands with very narrow 3-dB bandwidths are observed on the spectra. Due to the limited resolution (0.01 nm) of the optical spectrum analyzer (OSA, Ando AQ 6317B), the true 3-dB bandwidths of the two transmission bands cannot be resolved. They are estimated to be 3.25 pm (peak 1 at 1542.257 nm) and 2.85 pm (peak 2 at 1542.403 nm) from the simulation result. The wavelength spacing between the two channels is ~0.146 nm.

## 4. Application of the IGAFBG in Dual Wavelength Lasing

The IGAFBG can be used as a dual-channel filter for dual-wavelength fiber laser emission. To demonstrate it, we constructed a linear cavity fiber laser as schematically shown in Fig. 4. The laser cavity consists of a uniform FBG which couples the pump into the cavity and simultaneously acts as one cavity mirror around 1542.3 nm, a 4.3-m-long erbium-doped fiber (EDF) as the gain medium, a polarization controller (PC) to fine tune the cavity birefringence, an IGAFBG as the dual-channel filter (its transmission spectrum is shown in Fig. 3), and a chirped FBG (CFBG) as another cavity mirror and the output coupler. All the used gratings (uniform FBG, IGAFBG and chirped FBG) are fabricated by using the same standard single-mode hydrogen-loaded fibers. The uniform FBG has a length of 1 cm with 33-dB reflectivity, 0.22 nm 3-dB bandwidth, and a Bragg wavelength at 1542.33 nm. This uniform FBG filters out and reflects the two transmission bands coming from the IGAFBG back into the laser cavity. The chirped FBG has a length of 3 cm, whose Bragg wavelength is centered at 1545.5 nm (chirp rate is 4.5 nm/cm) with a stopband ~18 nm in the transmission mode, and a reflectivity of ~94%. By adjusting the polarization controller carefully, an output spectrum of the laser is shown in Fig. 5(a), measured under a pump power of 300 mW. The laser emits simultaneously at 1542.257 nm and 1542.403 nm (wavelength separation is 0.146 nm), which exactly match the two pass bands shown in Fig. 3. The optical signal-to-noise ratio (OSNR) for these two lasing lines is around 45 dB. Fig. 5(b) shows the repeated scans of the two ultra-narrow lasing lines at two-minute intervals over half an hour at room temperature. The dual-wavelength lasing parameters, including central wavelength, wavelength spacing, 3-dB spectral bandwidths and OSNR, are all kept reasonably unchanged, indicating the good stability of the lasing output. A change in the temperature of the laser might shift the two lasing lines simultaneously since the reflection spectrum of the uniform FBG, IGAFBG and CFBG are temperature sensitive, and will drift by the same amount, but this will not affect the wavelength spacing and stability.

## 5. Discussion

Erdogan classified the types of FBGs into five main categories: uniform, apodized, chirped, phase-shifted and superstructure gratings, and described by Fig. 2 in [5]. It can be seen from Fig. 1(b), the proposed IGAFBG cannot be categorized into any of the five traditional FBG structures. It seems that the IGAFBG is a special case of the apodized FBG, since we use the inverse-Gaussian apodization function. However, "apodization" is defined as the suppression of the sidelobes in the reflection spectrum [17], by periodically increasing the modulation depth from the two ends to the center of the grating. Obviously, the IGAFBG enhances the sidelobes instead of suppressing them in the reflection spectrum and thus it does not belong to the apodized FBG.

It is well known that a FBG pair can form two transmission peaks [11]. Fig. 6 shows the variation of the effective index along the fiber axis of a FBG pair. For illustrative purposes as well, the size of the grating period is greatly exaggerated relative to the grating length. By comparing Fig. 6 and Fig. 1(b), it is clearly noted that the IGAFBG also distinguishes with the FBG pair, since an IGAFBG has no gaps and it has continually varying effective refractive index.

The separation of the two transmission bands of an IGAFBG can be flexibly changed by changing the grating length and the modulation depth. Two transmission peaks with spacing of 0.1 nm was successfully achieved in experiment, as shown in Fig. 7. To fabricate this 18-mm-long IGAFBG, we used 60 mW Argon laser and the same scanning speed distribution $V(z)$ as the one used in Fig. 3. In simulation, $n_{eff}$ is 1.447, $L$ is 18 mm, $\Lambda$ is 532.85 nm, and $\overline{\delta n_{eff}}$ is $2.3 \times 10^{-4}$. Narrower spacing of the two transmission bands can be realized by increasing the grating length and decreasing the Argon laser power. We used 50 mW Argon laser to fabricate a 25-mm-long IGAFBG and thus to achieve 0.07 nm separation between the two pass bands, as shown in Fig. 8. In simulation, $n_{eff}$ is 1.447, $L$ is 25 mm, $\Lambda$ is 532.85 nm, and $\overline{\delta n_{eff}}$ is $1.4 \times 10^{-4}$. The small discrepancy between the experiment result (solid line) and the simulation result (dashed line) may be due to the unavoidable misalignment of the fiber relative to the phase mask.

## 6. Conclusions

We have proposed and fabricated a new FBG structure. It is shown that an inverse-Gaussian apodized FBG possesses easily controllable dual-wavelength transmission peaks, which can be used as a dual-channel bandpass filter. Incorporating the FBG filter in a linear cavity fiber laser, stable dual-wavelength emission with a wavelength separation of 0.146 nm has been demonstrated. The dual-wavelength fiber laser has the characteristics of simple structure, low cavity loss and low cost.

**Figure Captions:**

1. (a) Calculated reflection spectrum of the proposed IGAFBG, where gap 1 and 2 can form two high transmittivity pass bands in the corresponding transmission spectrum. (b) $n_{eff}$ variation along the fiber axis of the IGAFBG, the dashed line is an inverse-Gaussian distribution.

2. Calculated reflection spectrum of a traditional Gaussian apodized FBG. The parameters used are the same as the proposed IGAFBG, except the apodization function is $B(z) = \exp\left\{\left[-4(\ln 2)z^2\right]/(L/3)^2\right\}$.

3. Transmission spectra of the proposed IGAFBG with wavelength spacing of 0.146 nm (peak 1 at 1542.257 nm and peak 2 at 1542.403 nm). Solid line, measured spectrum; dashed line, calculated spectrum. In simulation, $n_{eff} = 1.447$, $L = 12$ mm, $\Lambda = 532.85$ nm and $\overline{\delta n_{eff}} = 3 \times 10^{-4}$. Fig. 1(a) is the corresponding calculated reflection spectrum of the same IGAFBG.

4. Schematic diagram of the dual-channel fiber laser with an IGAFBG filter in its linear all-fiber cavity. EDF, erbium doped fiber; PC, polarization controller; IGAFBG, inverse-Gaussian-apodized fiber Bragg grating; CFBG, chirped fiber Bragg grating; OSA, optical spectrum analyzer.

5. Output laser spectra. (a) two lasing lines at 1542.257 nm and 1542.403 nm (within 4.5 nm wavelength domain); (b) repeated scans of the two lasing lines at two-minute intervals over half an hour at room temperature (within 1 nm wavelength domain).

6. Effective refractive index variation along the fiber axis of a standard FBG pair.

7. Transmission spectra of an IGAFBG with wavelength spacing of 0.1 nm. Solid line, measured spectrum; dashed line, calculated spectrum. In simulation, $n_{eff}$ = 1.447, $L$ = 18 mm, $\Lambda$ = 532.85 nm and $\overline{\delta n_{eff}} = 2.3 \times 10^{-4}$.

8. Transmission spectra of an IGAFBG with wavelength spacing of 0.07 nm. Solid line, measured spectrum; dashed line, calculated spectrum. In simulation, $n_{eff}$ = 1.447, $L$ = 25 mm, $\Lambda$ = 532.85 nm and $\overline{\delta n_{eff}} = 1.4 \times 10^{-4}$.

**Figures**

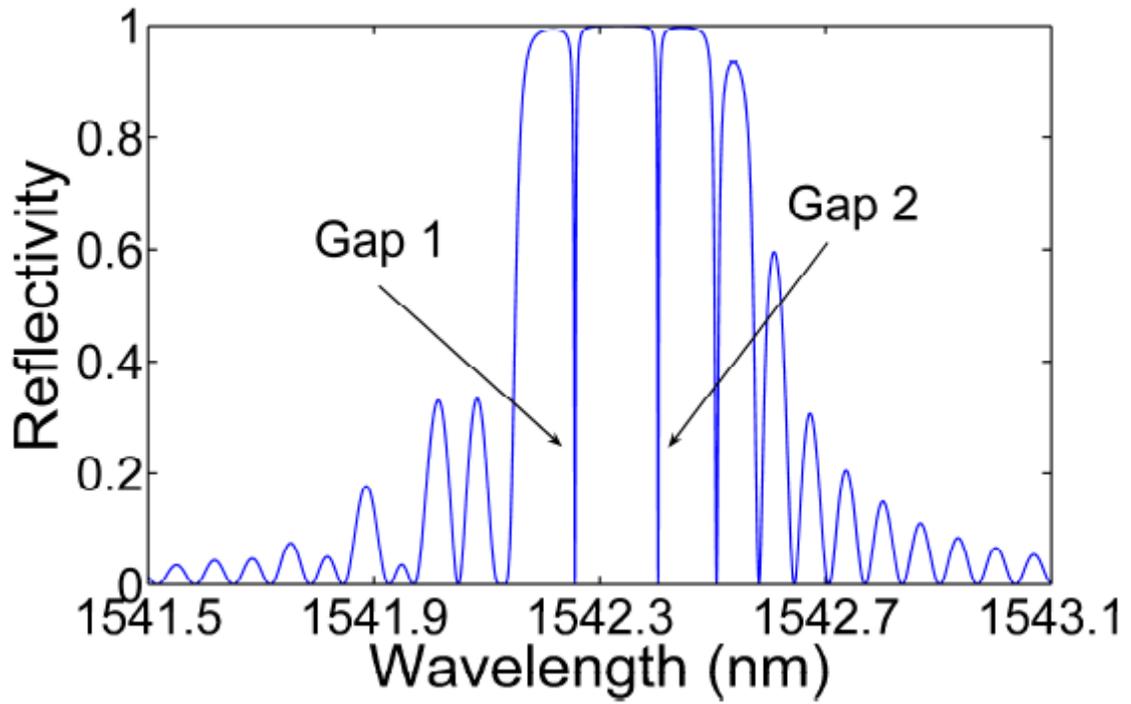

(a)

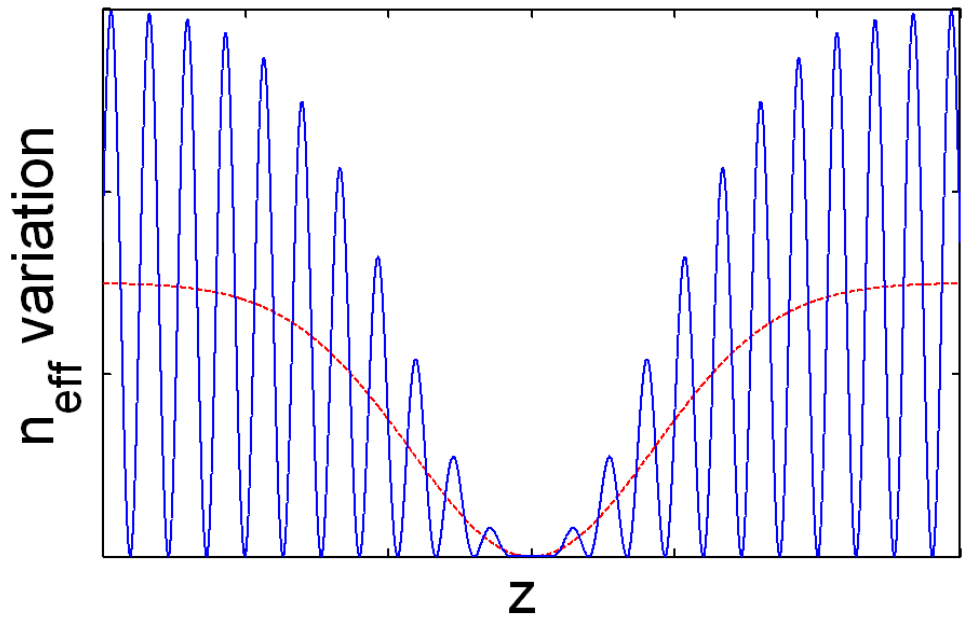

(b)

Fig. 1.
14

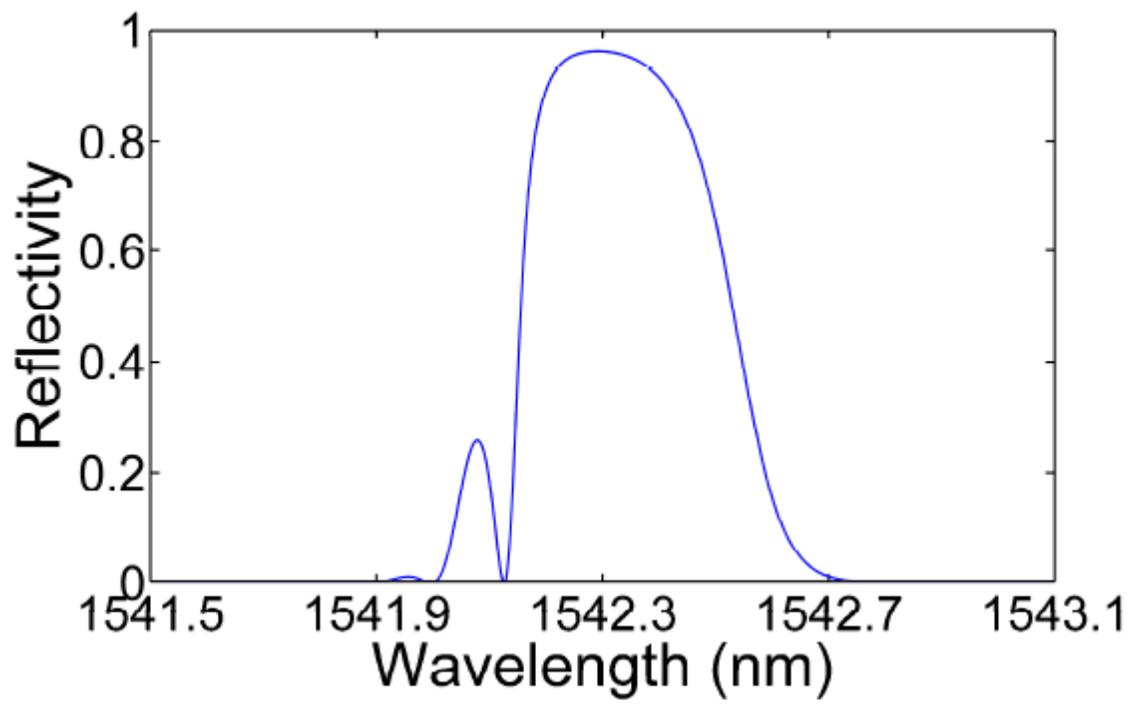

Fig. 2.

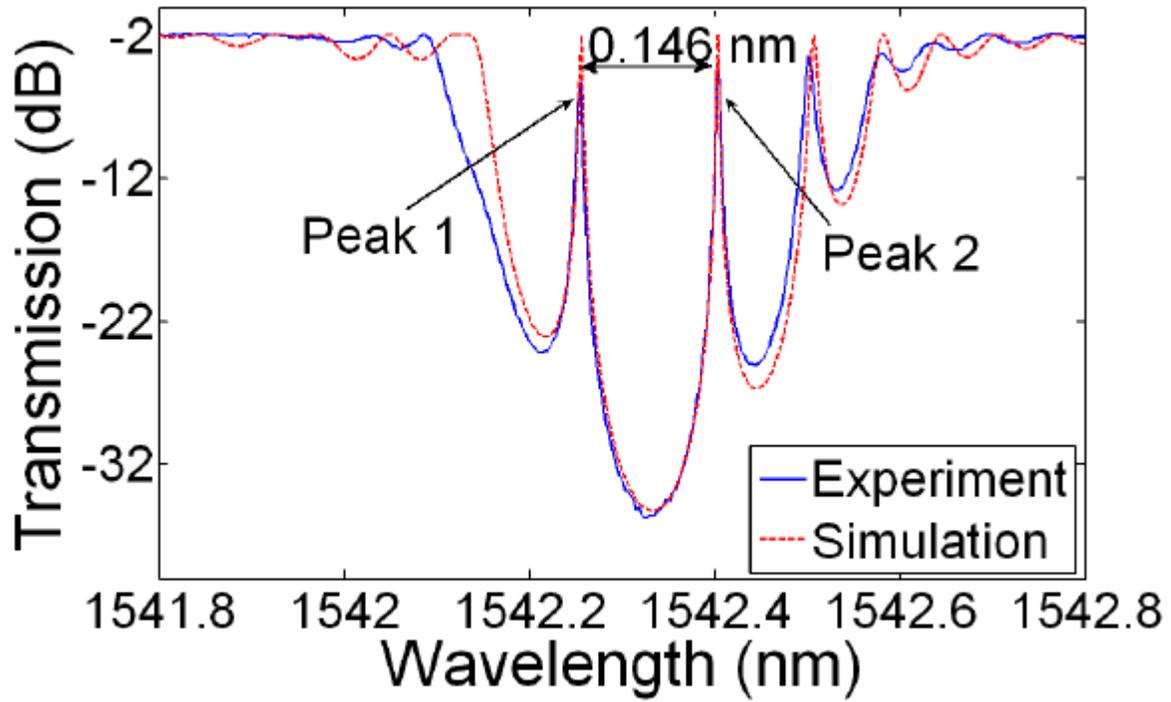

Fig. 3.

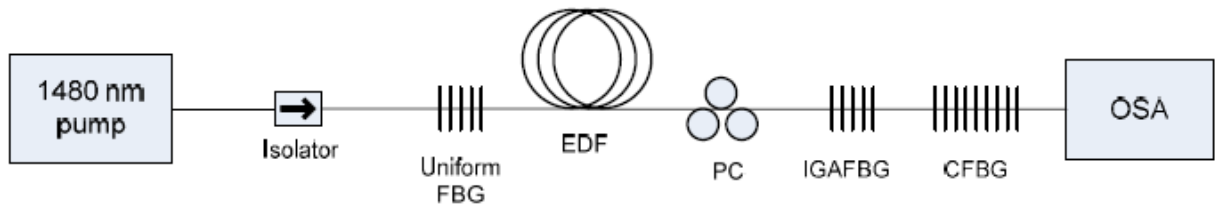

Fig. 4.

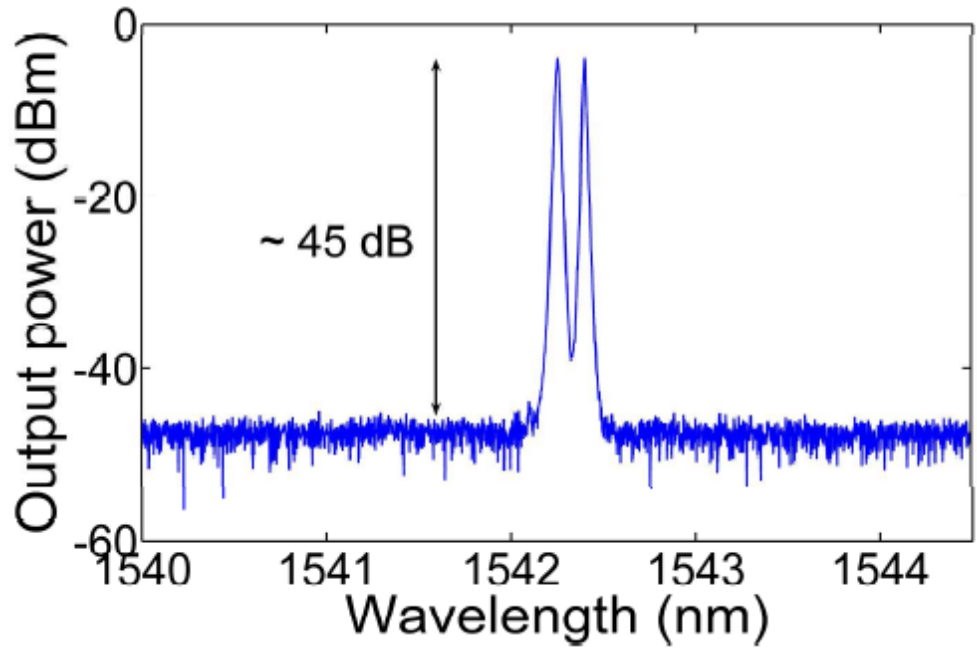

(a)

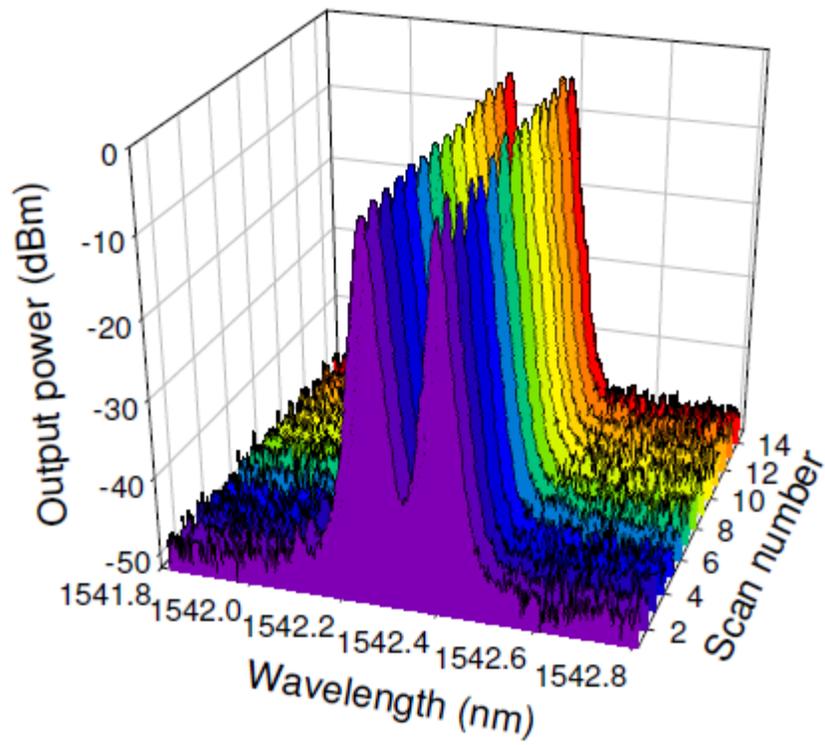

(b)

Fig. 5.

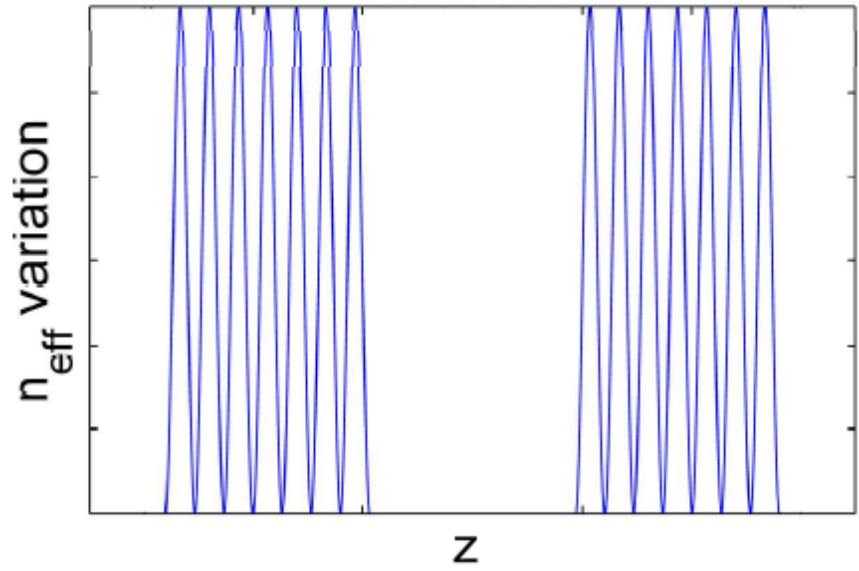

Fig. 6.

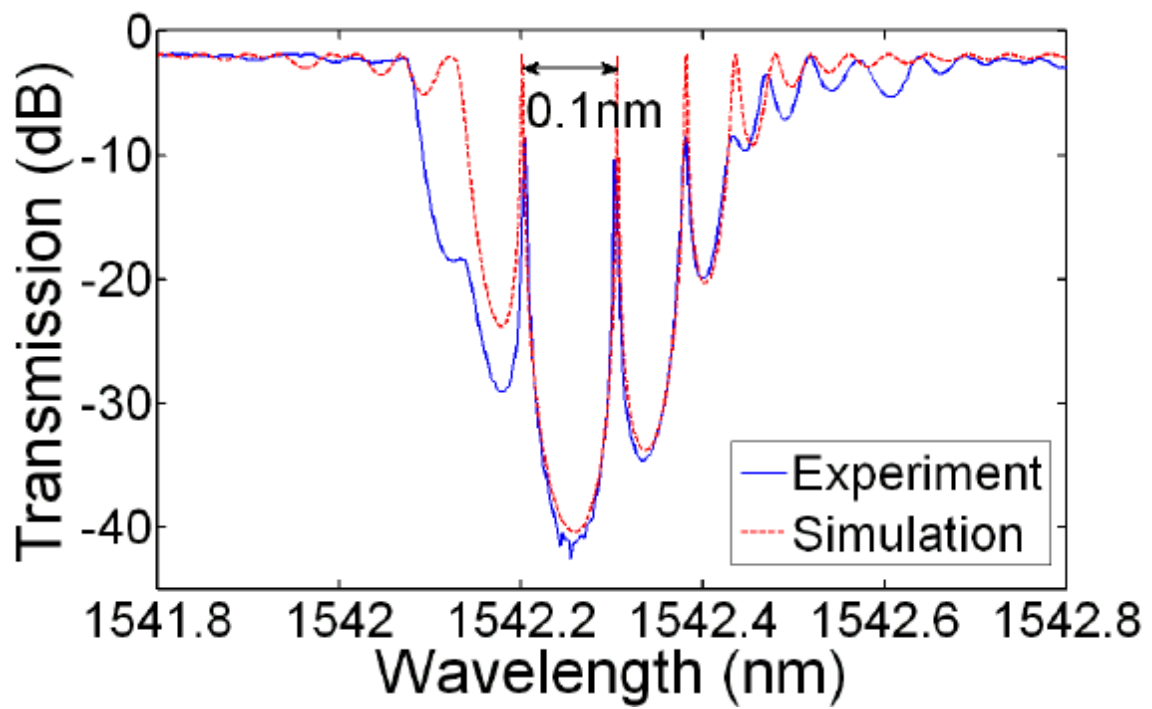

Fig. 7.

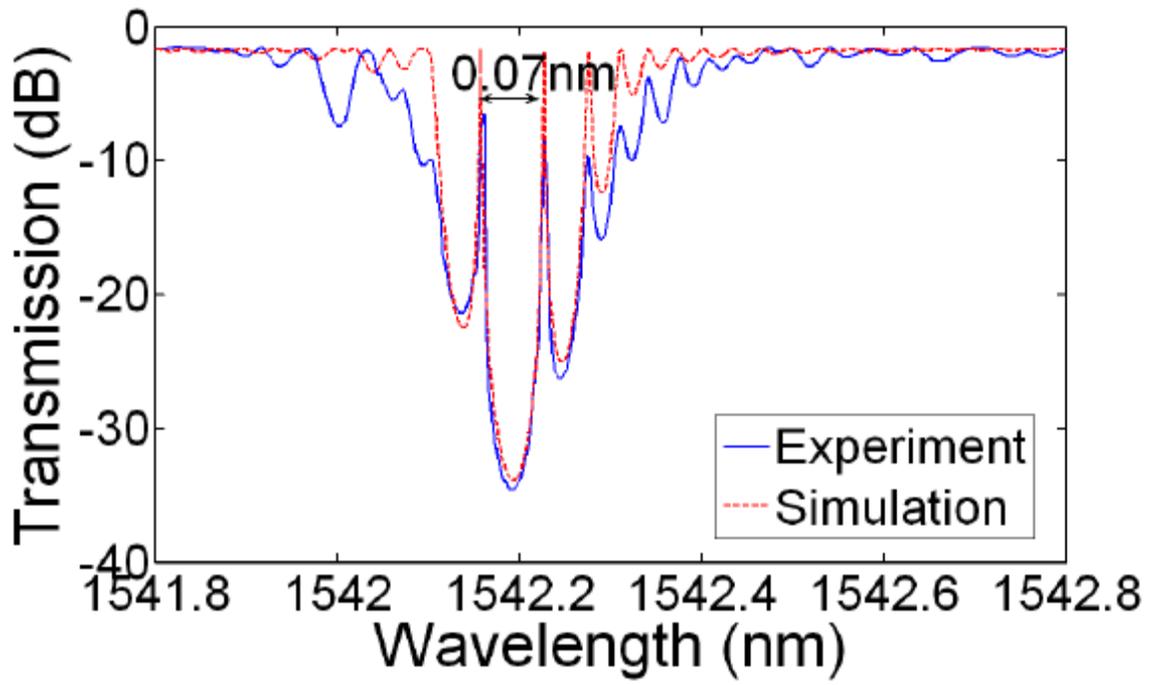

Fig. 8.